# Spin-orbit coupling induced Van Hove singularity in proximity to a Lifshitz transition in $Sr_4Ru_3O_{10}$

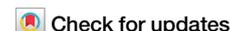
Check for updates

Carolina A. Marques ①[1,9,10] ✉, Philip A. E. Murgatroyd ①[1,10], Rosalba Fittipaldi[2], Weronika Osmolska ①[1], Brendan Edwards ①[1], Izidor Benedičič[1], Gesa-R. Siemann ①[1], Luke C. Rhodes ①[1], Sebastian Buchberger ①[1,3], Masahiro Naritsuka ①[1], Edgar Abarca-Morales[1,3], Daniel Halliday[1,4], Craig Polley[5], Mats Leandersson[5], Masafumi Horio ①[6], Johan Chang ①[6], Raja Arumugam[2], Mariateresa Lettieri ①[2], Veronica Granata ①[7], Antonio Vecchione ①[2], Phil D. C. King ①[1] ✉ & Peter Wahl ①[1,8]

Van Hove singularities (VHss) in the vicinity of the Fermi energy often play a dramatic role in the physics of strongly correlated electron materials. The divergence of the density of states generated by VHss can trigger the emergence of phases such as superconductivity, ferromagnetism, metamagnetism, and density wave orders. A detailed understanding of the electronic structure of these VHss is therefore essential for an accurate description of such instabilities. Here, we study the low-energy electronic structure of the trilayer strontium ruthenate $Sr_4Ru_3O_{10}$, identifying a rich hierarchy of VHss using angle-resolved photoemission spectroscopy and millikelvin scanning tunneling microscopy. Comparison of k-resolved electron spectroscopy and quasiparticle interference allows us to determine the structure of the VHss and demonstrate the crucial role of spin-orbit coupling in shaping them. We use this to develop a minimal model from which we identify a mechanism for driving a field-induced Lifshitz transition in ferromagnetic metals.

Van Hove singularities (VHss) in the density of states appear due to stationary points, $\nabla E(\mathbf{k}) = 0$, in the quasiparticle dispersion relation $E(\mathbf{k})$[1]. They are a direct consequence of the periodicity of the crystal lattice, appearing naturally at high symmetry points in the Brillouin zone (BZ), but also away from these points due to higher-order hopping terms and band hybridizations. The consequences of VHss for the properties of a material are intricately linked to the abrupt change or divergence in density of states associated with it. In two-dimensional systems, band minima or maxima lead to a step change in the density of states while saddle points result in logarithmic or higher order divergences dependent on their symmetry and parameters of the band structure[2]. Tuning such a density of states divergence through the Fermi energy $E_F$ can be expected to drive electronic instabilities, concomitant with the resulting Lifshitz transition where the topology of the Fermi surface changes. Indeed, many properties of strongly correlated electron materials have been associated with VHss and accompanying Lifshitz transitions, including, e.g., metamagnetic transitions in heavy fermion systems[3–5], the pseudogap phase[6–8] and high-$T_c$ superconductivity[9] in cuprates, the interplay between superconducting and Mott insulating states in twisted bilayer graphene[10–13] and charge density wave formation and superconductivity in kagome materials[14,15].

There is thus a need to identify model systems through which to develop a coherent understanding of the role of VHss in shaping the collective states of quantum materials. A particularly promising system is the Ruddlesden-Popper series of strontium ruthenates $Sr_{n+1}Ru_nO_{3n+1}$, which are known to exhibit multiple VHss close to the Fermi energy[16,17]. While the $n \to \infty$ member, $SrRuO_3$, is a ferromagnet[18], the first two members with $n = 1$ and $n = 2$ are nonmagnetic. $Sr_2RuO_4$ ($n = 1$) hosts unconventional superconductivity which can be markedly enhanced by uniaxial pressure[19], concomitant with a VHs being driven through the Fermi level[20]. $Sr_3Ru_2O_7$ ($n = 2$) is on the verge of magnetism, exhibiting strong ferromagnetic fluctuations close to criticality and a metamagnetic transition in applied magnetic field[21]. Unusual scaling relations in its thermodynamic properties have, in turn, been linked to the presence of higher-order VHss near the Fermi level[17]. The $n = 3$ member, $Sr_4Ru_3O_{10}$, is the first member in which a bulk ferromagnetic ground state is realized[22]. Its crystal structure consists of trilayers of $SrRuO_3$ stacked along the $c$-axis, connected by the apical oxygen of the $RuO_6$ octahedra (Fig. 1a). These $RuO_6$ octahedra are furthermore rotated around the $c$-axis, leading to an orthorhombic unit cell.

[1]SUPA, School of Physics and Astronomy, University of St Andrews, North Haugh, St Andrews KY16 9SS, UK. [2]CNR-SPIN, UOS Salerno, Via Giovanni Paolo II 132, Fisciano I-84084, Italy. [3]Max Planck Institute for the Chemical Physics of Solids, Nöthnitzer Straße 40, 01187 Dresden, Germany. [4]Diamond Light Source Ltd, Diamond House, Harwell Science and Innovation Campus, Didcot OX11 0DE, UK. [5]MAX IV Laboratory, Lund University, 221 00 Lund, Sweden. [6]Physik-Institut, Universität Zürich, Winterthurerstrasse 190, CH-8057 Zürich, Switzerland. [7]Dipartimento di Fisica "E. R. Caianiello", Università di Salerno, I-84084 Fisciano, Salerno, Italy. [8]Physikalisches Institut, Universität Bonn, Nussallee 12, 53115 Bonn, Germany. [9]Present address: Physik-Institut, Universität Zürich, Winterthurerstrasse 190, CH-8057 Zürich, Switzerland. [10]These authors contributed equally: Carolina A. Marques, Philip A. E. Murgatroyd.
✉e-mail: cdamarques@outlook.com; pdk6@st-andrews.ac.uk

$Sr_4Ru_3O_{10}$ undergoes a ferromagnetic transition with a Curie temperature $T_C = 105$ K (Fig. 1b, Supplementary Fig. 1), with the magnetization aligned parallel to the $c$-axis[22]. At $T^* = 50$ K, a secondary peak in the magnetic susceptibility is observed which is associated with a metamagnetic transition triggered by in-plane magnetic field[22,23], and accompanied by an increase in the volume of the unit cell[24,25]. For magnetic fields applied in the $ab$-plane, $T^*$ decreases, reaching 0 K at ~2.5 T[26]. A link between the metamagnetic properties and VHss in proximity of the Fermi energy has been postulated[27], however direct experimental evidence for this is so far lacking[28–30].

Here, in a combined study of $Sr_4Ru_3O_{10}$ by angle-resolved photoemission spectroscopy (ARPES) and scanning tunneling microscopy and spectroscopy (STM/STS), we provide a comprehensive picture of the low-energy electronic structure, identifying multiple VHss in the vicinity of the Fermi level. Relating the spectral function measured by ARPES and quasiparticle interference (QPI) reveals that the VHs closest to the Fermi energy emerges due to a combined influence of octahedral rotations and spin-orbit coupling. Our results suggest a mechanism for the metamagnetic transition where the magnetization direction in conjunction with spin-orbit coupling drives a field-induced Lifshitz transition.

## Results
### Surface electronic structure
Due to the layered structure of the strontium ruthenates, cleavage results in atomically clean and flat SrO-terminated surfaces[29,31–35]. Figure 1c shows such a SrO surface of $Sr_4Ru_3O_{10}$. We find large terraces with only a few defects. Two types of defects centred in the hollow site between Sr atoms can be seen in Fig. 1c: defects with a $C_4$ symmetry that we associate with substitutional Ru site defects, and cross-like defects with $C_2$ symmetry. The substitutional Ru site defects occur with two different chiralities due to the octahedral rotations. The cross-like defects can be attributed to $CO_2$ complexes resulting from the chemisorption of CO molecules at the surface[36]. Both types of defects act as scattering centers, giving a strong signal for QPI which we discuss below. From topographies with lateral sizes larger than 100 nm, we determine a density of surface defects of ~0.78%. The inset of Fig. 1c shows a magnified view, where the square atomic lattice centred on the Sr atoms is visible. Density functional theory (DFT) calculations suggest that the oxygen octahedra in the surface layer exhibit a larger octahedral rotation angle compared to the bulk, close to 11° (Supplementary Fig. 2), although not resulting in additional periodicities. Consistent with the calculations, we do not see any evidence for a surface reconstruction.

An overview of the electronic structure, as obtained by ARPES, is shown in Fig. 1d. The data indicates a complex multi-band fermiology in this compound, consistent with other reports[28,30]. Strong matrix element variations are found throughout the tetragonal Brillouin zone. Large nearly square-shaped electron pockets are visible around the Brillouin zone center, whose corners reach approximately to the orthorhombic Brillouin zone boundary along the $(0, 0) - (\pi, \pi)$ direction ($\Gamma-X$) (Supplementary Fig. 3).

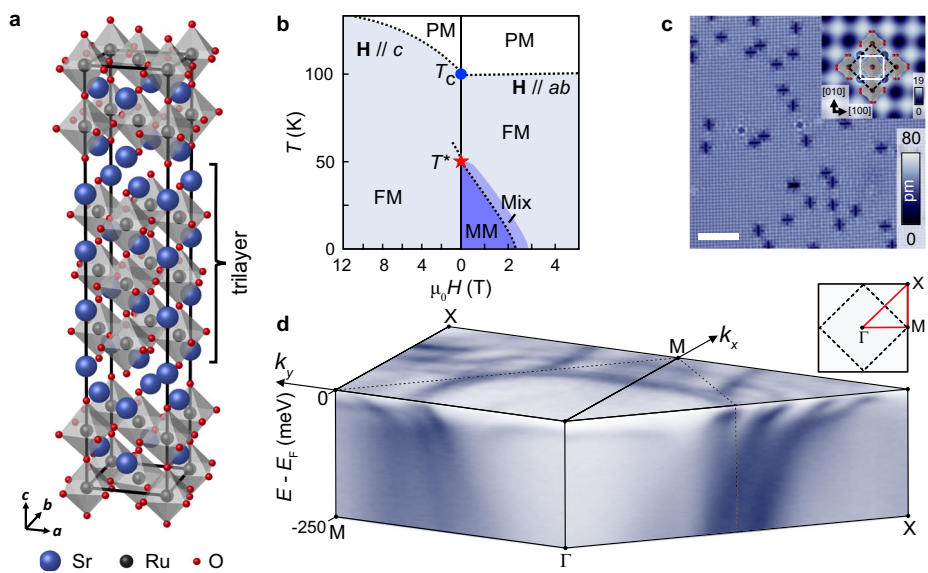

**Fig. 1 | Crystal structure, magnetic phase diagram, surface morphology and electronic structure of $Sr_4Ru_3O_{10}$. a** Crystal structure of $Sr_4Ru_3O_{10}$ in the orthorhombic unit cell[22]. **b** Phase diagram of $Sr_4Ru_3O_{10}$ for a magnetic field applied along the $c$-axis (left) and in the $a$-$b$ plane (right), based on our magnetization measurements in Supplementary Fig. 1 and refs. 24,26. The dotted line indicates the paramagnetic (PM) to ferromagnetic (FM) transition with $T_C \approx 100$ K at 0 T. There is an increase in magnetic susceptibility at $T^* = 50$ K (red star), identified as a metamagnetic transition. $T^*$ increases for magnetic fields up to 1 T parallel to the $c$-axis. For a magnetic field applied in the $a$-$b$ plane, $T^*$ decreases, reaching 0 K at ~ 2.5 T, enclosing a metamagnetic phase (MM) separated from the FM phase by a mixed phase region (Mix). **c** Typical topography of the surface of $Sr_4Ru_3O_{10}$, showing point defects with two distinct chiralities ($V_{set} = -5$ mV, $I_{set} = 91$ pA, $T = 75$ mK, scale bar: 5 nm). The inset shows a close-up over 4 × 4 unit cells with the atomic lattice superimposed ($V_{set} = 10$ mV, $I_{set} = 500$ pA). The dashed square indicates the orthorhombic unit cell, and the solid square the tetragonal unit cell. Crystallographic notation throughout this work refers to the tetragonal unit cell. **d** Fermi surface map for $Sr_4Ru_3O_{10}$ determined by angle-resolved photoemission ($T = 20$ K, $h\nu = 67$ eV, linear-vertical polarization), integrated within $E_F \pm 20$ meV. Corresponding $E$ vs. $\mathbf{k}$ dispersions measured along the $\Gamma - M$ and $\Gamma - X$ high-symmetry directions, providing an overview of the underlying electronic band structure, measured under the same experimental conditions as the Fermi surface. Schematics of the tetragonal (solid square) and orthorhombic (dashed square) Brillouin zones are shown in the top right corner.

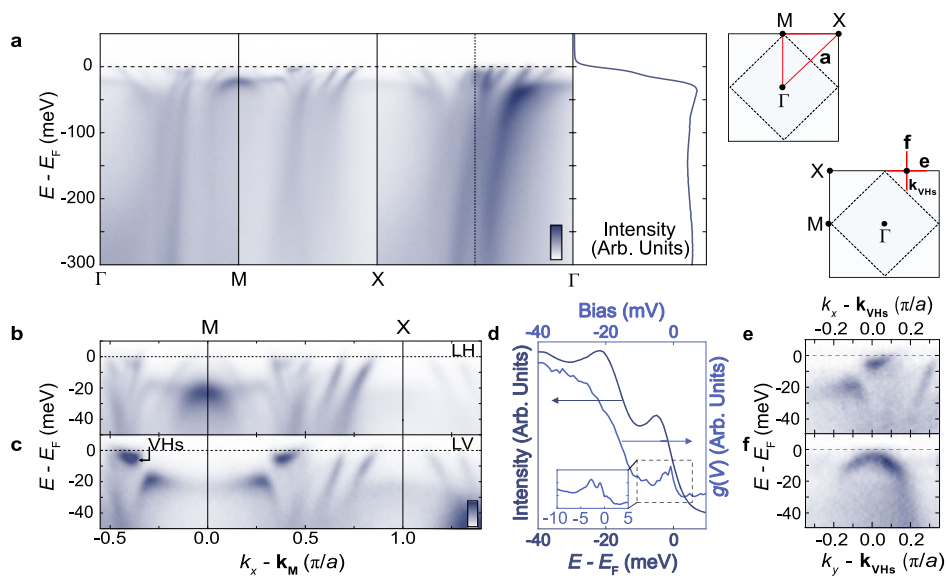

**Fig. 2 | Low-energy electronic structure of $Sr_4Ru_3O_{10}$. a** Dispersions measured by ARPES along $\Gamma - M - X - \Gamma$ [$T = 20$ K, $h\nu = 27$ eV ($\Gamma - M$, $X - \Gamma$) and $h\nu = 32$ eV (M − X); the data are shown as the sum of measurements performed using linear horizontal (LH) and linear vertical (LV) polarized light – see Supplementary Fig. 4 for the raw data]. The integrated intensity is shown on the right-hand side. A rich low-energy electronic structure with multiple VHss can be seen within 50 meV of the Fermi energy at the $\Gamma$ and M points. High-resolution measurements in the vicinity of the Fermi energy measured along M−X for LH, (**b**), and LV, (**c**) polarized light ($h\nu = 32$ eV). For LH polarization, the VHs at the M point is clearly visible, whereas LV polarization more clearly highlights hybridization gaps between M and X. **d** Comparison of the **k**-integrated intensity of the spectral function along M−X shown in (**c**) to a tunneling spectrum, $g(V)$, obtained at 0 T and $T = 75$ mK. Two broad peaks in the integrated intensity obtained from ARPES are observed at ~ 5 meV and ~20 meV below the Fermi level. The enhanced energy resolution of tunneling spectroscopy reveals that the lower binding energy peak is composed of multiple peaks in the bias voltage range of −5 to 0 mV (see inset) ($V_{set} = 40$mV, $I_{set} = 500$ pA, $V_L = 0.4$ mV; Inset: $V_{set} = 10$ mV, $I_{set} = 500$ pA, $V_L = 0.2$ mV). ARPES dispersions measured along the, (**e**) M-X, and (**f**) orthogonal direction in the vicinity of the identified VHs, as extracted from our Fermi surface mapping data as shown in Supplementary Fig. 5. A characteristic saddle point character of the dispersions is evident, with electron- and hole-like bands along the M-X and orthogonal directions, respectively. The insets in the top right show the relevant **k**-space paths for the ARPES measurements in (**a, e, f**).

These are in good qualitative agreement with calculations of the expected spin-minority Fermi surfaces[29,37]. Around the corners of the tetragonal Brillouin zone, we observe a complex set of additional intertwined Fermi pockets. These have small matrix elements close to the tetragonal Brillouin zone center, although replicas of these states are also visible close to the Brillouin zone center at selected photon energies (see, e.g., Supplementary Fig. 4).

To investigate the electronic structure from which this complex fermiology derives in detail, we show in Fig. 2a dispersions measured along the high-symmetry $\Gamma - M - X - \Gamma$ path of the tetragonal Brillouin zone. Consistent with the large number of Fermi pockets, we find a complex multi-band electronic structure. Sharp quasiparticles are visible within ≈ 30 meV of the Fermi level, but the spectral features quickly broaden with increasing binding energy, and the measured states show kink-like features around this energy scale, with a reduction in quasiparticle velocity at the Fermi level. A similar phenomenology has been observed in the single-layer compound[38], where the associated deviations in linearity of the real part of the self-energy calculated by dynamical mean-field theory have been attributed to a crossover from a Fermi liquid to a more incoherent regime. In this respect, we note that $Sr_4Ru_3O_{10}$ is known to host a Fermi liquid ground state, with a $T^2$ temperature dependence of its resistivity which persists up to comparable energy scales as in $Sr_2RuO_4$[39].

**Low-energy electronic structure and hybridization gaps**
Beyond this conceptual similarity, however, we note that there are significantly more low-energy states within ~50 meV of the Fermi level in $Sr_4Ru_3O_{10}$ than in $Sr_2RuO_4$. Notably, at an energy scale of $E - E_F \approx -20 - 30$ meV, we find several rather flat band features that contribute a high density of states near the Fermi level (Fig. 2a). The state at $\Gamma$ at a binding energy of ~25 meV has previously been observed, and assigned as part of the spin-majority states from spin-resolved photoemission[30]. In addition, our measurements reveal a rich hierarchy of VHss around the M point at the Brillouin zone face, shown in more detail in Fig. 2b, c.

At $E - E_F = \sim -20$ meV at the M-point, the top of a hole band and bottom of a weakly-dispersing electron band intersect (Fig. 2b). This reflects the folding of two VHss onto each other due to the octahedral rotations in this structure, which render the $\Gamma - M$ and $M - X$ directions equivalent within the orthorhombic Brillouin zone. Interestingly, a clear hybridization gap develops along the electron-like part of its dispersion, located at a momentum of $k_x \sim \pm 0.3 \frac{\pi}{a}$. This is particularly evident when measured using linear-vertical polarization (Fig. 2c).

These features are also evident in k-integrated measurements of the ARPES intensity of Fig. 2c, plotted in Fig. 2d. Two peaks are resolved, one at ~20 meV below the Fermi level and one which is peaked at ~5 meV below $E_F$, being cut off by the Fermi function. They correspond to the high-intensity hybridized/gapped states seen in Fig. 2c either side of M, with very similar EDCs found if integrating over a smaller momentum range close to these band features. We show the **k**-integrated intensity and high-resolution tunneling spectra measured at 75mK in Fig. 2d, for a comparison of the effective density of states probed by both techniques. The two-peak structure evident in the ARPES measurements is also visible in STM, with finite density of states close to the Fermi level and a second broad hump visible in the STM at a binding energy of around 20 meV. However, the higher resolution of the tunneling spectrum reveals that the lower binding energy feature (the single peak observed in ARPES close to the Fermi level) is in fact composed of multiple contributions, with a well-defined two-peak structure observed in STM within a bias voltage range of −5 to 0 mV, shown in the inset.

The gap seen in ARPES (Fig. 2c) can be conceptually understood as emerging from the crossing of two bands, which hybridize to form a gap as illustrated in Fig. 3a. Through ARPES and QPI we show how this results in the formation of two new VHss in the band structure, one at the top and one at the bottom edge of the hybridization gap (orange lines). The uppermost Van Hove singularity is extremely close to the Fermi energy and should therefore be the most relevant VHs for the metamagnetic properties of $Sr_4Ru_3O_{10}$. This VHs does not exist at a high-symmetry point within the Brillouin zone, instead being found at approximately $\left(0, 0.62\frac{\pi}{a}\right)$ and symmetry-equivalent points (Supplementary Fig. 5a). While the band disperses upwards along the M-X high-symmetry direction shown in Fig. 2c, e, we show in Fig. 2f and Supplementary Fig. 5 that there is a hole-like dispersion in the orthogonal direction, classifying this as a saddle point in the electronic structure. To further study this on the relevant low energy scales close to $E_F$, we perform QPI measurements within ± 5 mV around the Fermi level. Figure 3b shows the Fourier transform $\tilde{g}(V, \mathbf{q})$ of a $g(V, \mathbf{r})$ map in zero magnetic field measured just below the Fermi level at $V = -1.4$ mV. Several characteristic scattering vectors are observed that are consistent with previous work[29]. Specifically, the three rings centered at $\mathbf{q} = 0$ with $q \sim 1/2$ can be related to bands of minority-spin character[29,37]. We extract the positions of two of these rings, red and green circles, corresponding to intra-band scattering, while the middle ring corresponds to inter-band scattering between the two. Converting these scattering vectors (indicated by arrows in Fig. 3b) into $k$-space (Fig. 3e), we find excellent agreement with the large electron Fermi surfaces observed in our ARPES measurements. We also identify two additional sharp scattering vectors in Fig. 3b, whose positions are shown by blue and orange circles. When transforming their positions into $k$-space, we find that they match the tips of the leaves of the clover-shaped Fermi pocket centered at X. The scattering vectors in $k$-space corresponding to the QPI patterns identified in Fig. 3b are shown as arrows with the corresponding colors in Fig. 3e, and yield Fermi contours which are again in excellent agreement with our ARPES measurements.

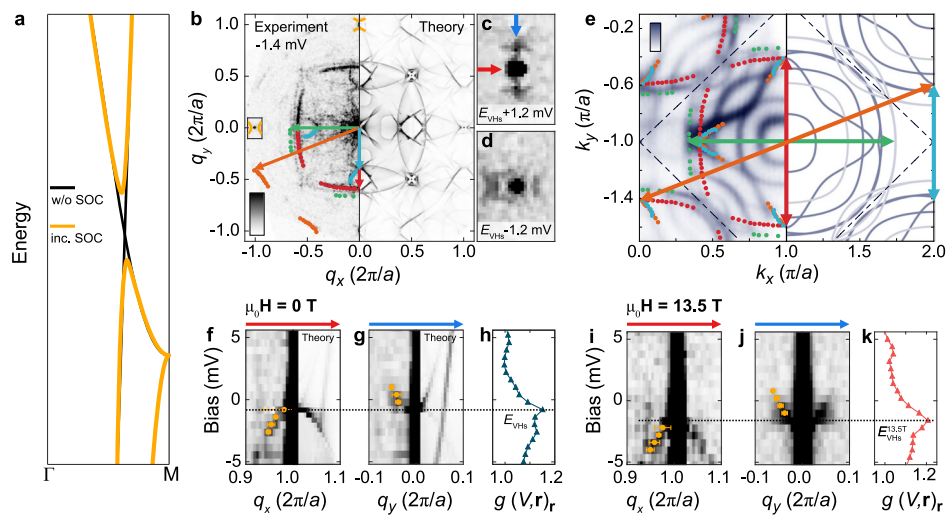

**Fig. 3 | Detailed structure of the Van Hove singularity closest to the Fermi level. a** Illustration of two bands crossing (black lines) that form a gap due to spin-orbit coupling (yellow). The top and bottom edges of the gap form VHss. **b** Fourier transformation $\tilde{g}(\mathbf{q}, V)$ (symmetrized, see also Supplementary Fig. 6) of a $g(\mathbf{r}, V)$ map at $V = -1.4$ mV and $\mu_0 H = 0$ T. Left half shows the experimental data, right half the QPI obtained from a cLDOS calculation (see methods for details). The colored circles are peak positions extracted from fits to the QPI patterns at −0.2 mV. The arrows indicate the corresponding scattering vectors. **c, d** show close-ups around the (1,0) atomic peak, indicated by the black box in (**b**), at 1.2 mV above and below $E_{\text{VHs}}$, respectively. **e** ARPES Fermi surface (left, $h\nu = 32$ eV, centred at the X-point of the tetragonal Brillouin zone ($\Gamma_{10}$ of the orthorhombic zone), $T = 20$ K, integrated over $E_F \pm 5$ meV) with comparison to a minimal model of the electronic structure (right). The ARPES data is shown as the sum of Fermi surface maps measured with LH and LV polarized light (Supplementary Fig. 7). The light blue lines on the right panel correspond to a replica of the minimal model shifted by 40 meV to account for the multilayer nature of the material, as discussed for Fig. 4. The symbols are the points extracted from the QPI measurements, shown in (**b**), highlighting excellent agreement between the two techniques. The arrows are the scattering vectors corresponding to the QPI symbols of the same color and as shown in (**b**. **f, g**), Line cuts through $\tilde{g}(\mathbf{q}, V)$ across the (1, 0) atomic peak, left half experimental data, right half cLDOS calculation. **f** Cut along the [1, 0] direction (red arrow in **c**) and, (**g**), along the [0, 1] direction (blue arrow in **c**). **h** Spatially-averaged spectrum $\langle g(V, \mathbf{r}) \rangle_\mathbf{r}$, with the energy of $E_{\text{VHs}}$ indicated ($V_{\text{set}} = 10$ mV, $I_{\text{set}} = 450$ pA, $V_L = 0.6$ mV, $T = 75$ mK). **i–k** show the same cuts as (**f, g**) and average spectrum as (**h**), but measured in a magnetic field of $\mu_0 H = 13.5$ T. The same saddle point VHs can be seen as in zero field, however, shifted down by about 1 meV, confirming a dominant spin-majority character.

Building on this agreement, we now turn to identifying signatures of the near-$E_F$ VHs evident in the ARPES measurements. Signatures of VHss in QPI are expected to appear as distinct scattering patterns near $\mathbf{q} = 0$ and close to the atomic peaks[40]. Indeed, we observe a set of QPI features close to the atomic peaks at (0, 1) and (1, 0) (yellow arcs in Fig. 3b). Close-ups around the atomic peak at (1, 0) (Fig. 3c, d) reveal a change in the orientation of the dominant scattering pattern between $E_{\text{VHs}} + 1.2$ mV, Fig. 3c, and $E_{\text{VHs}} - 1.2$ mV, Fig. 3d. Line cuts along the [1, 0] (red arrow) and [0, 1] (blue arrow) directions (Fig. 3f, g) exhibit a hole-like dispersion along [1, 0] with a maximum at −0.8 mV, and an electron-like dispersion with a minimum at the same energy along [0, 1]. The change in sign of the curvature between the [1, 0] and [0, 1] directions allows us to identify $E_{\text{VHs}}$ with a saddle point VHs, entirely consistent with our ARPES data. In agreement with this picture, the spatially-averaged spectrum $\langle g(V, \mathbf{r}) \rangle_\mathbf{r}$ (Fig. 3h) shows a peak at $E_{\text{VHs}}$, the same energy at which the dispersions collapse onto the atomic peak. When applying an out-of-plane magnetic field of 13.5 T, we find that the observed dispersion of the VHs moves to lower energies (Fig. 3i–k) by ≈ 1 meV, behavior that is indicative of the states being of spin-majority character.

To connect the observation of the hybridization gap close to the Fermi level by ARPES with the saddle-point VHs identified in QPI, we develop a minimal model. The two-dimensional nature of the electronic structure, where the dispersion in the $k_z$ direction is practically negligible, means that it is in principle sufficient to consider a single trilayer of $Sr_4Ru_3O_{10}$. Such a layer would still require 36 bands to fully describe its electronic structure. However, the similarity of the electronic structure to $Sr_2RuO_4$ and the extreme surface sensitivity of QPI allow us to consider only the top-most $RuO_2$ layer, enabling a description of the surface electronic structure in terms of a ferromagnetic monolayer of $Sr_2RuO_4$. We start with a model for the band structure of $Sr_2RuO_4$ with 11° octahedral rotation, as guided by our DFT calculations. We introduce ferromagnetism along the $c$-axis and include spin-orbit coupling (SOC), matching the relevant parameters to the band structure obtained from ARPES and STM. This captures key features of the electronic structure close to $E_F$, within ± 50 meV, in particular reproducing the saddle point VHs at M, and yielding a hybridization gap that opens away from the M point when SOC is included (Fig. 3a). The

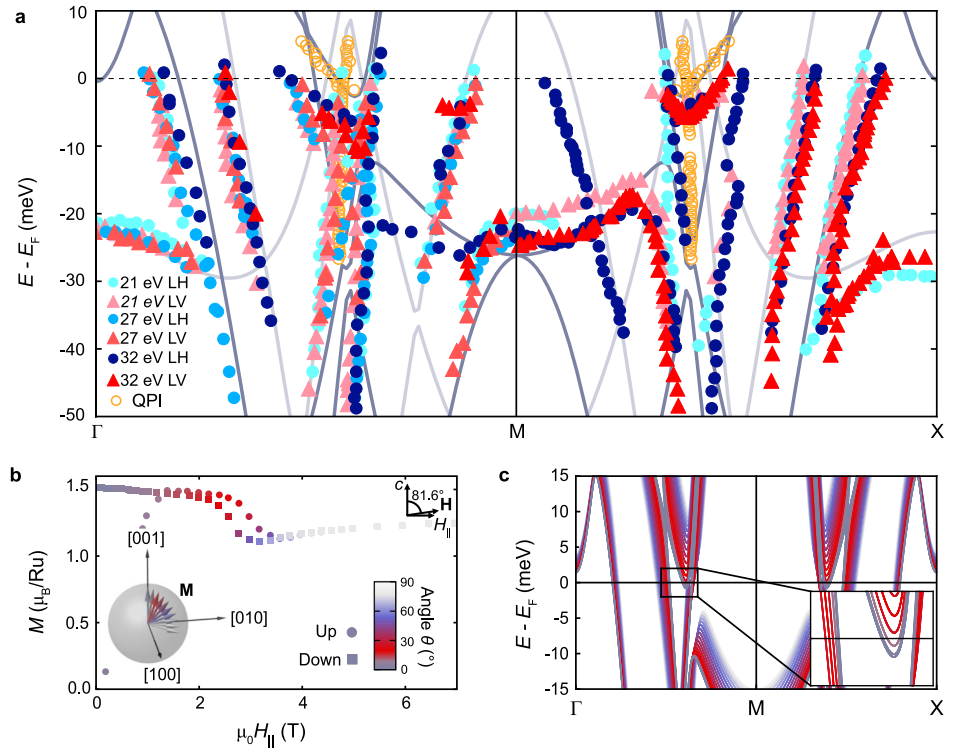

**Fig. 4 | Comprehensive picture of electronic structure and spin-orbit-coupling-driven Lifshitz transition. a** Comparison of the minimal model for the low-energy electronic structure with spin-orbit coupling to the electronic structure extracted from ARPES and QPI, showing excellent overall agreement (dark lines—model as fit to the QPI data; light lines - additional bands arising due to the multilayer nature of the material; colored points—dispersion extracted from ARPES (using the photon energies and polarizations indicated in the legend, see also Supplementary Fig. 10) and from QPI (Supplementary Fig. 11). **b** Magnetization **M** of bulk $Sr_4Ru_3O_{10}$ as a function of the in-plane component $\mu_0 H_\parallel$ of a field $\mu_0 \mathbf{H}$ applied at an angle of 81.6° from the c-axis (data reproduced from Fig. 2 of ref. [47]). Round symbols show the magnetization during the up-sweep starting from a zero-field cooled state, and square symbols when ramping the field down again. The metamagnetic transition can be seen at an in-plane field of about 2.8 T. The angle of the magnetization with respect to the c-axis is encoded in the color and illustrated in the inset in the lower left corner. **c** Band structure of the minimal model for magnetization along the [001]-axis ($\theta = 0°$) and with increasing angle $\theta$ towards the in-plane [110]-direction ($\theta = 90°$). The axes are relative to the tetragonal unit cell as in Fig. 1c. The inset shows a close-up around the VHs, showing in more detail the curves for angles from $\theta = 0°$ to $\theta = 25°$.

Fermi surface obtained from the model is consistent with QPI and with many of the pockets observed by ARPES (dark solid lines in Fig. 3e). These features are also captured in DFT calculations of bulk ferromagnetic $Sr_4Ru_3O_{10}$[37], which display qualitative similarities with our broad-scale measured dispersions in ARPES (Fig. 2). However, the location of the VHss depends sensitively on details of the DFT calculation, where exchange splitting is typically overestimated[30], and disentangling the correct low-energy electronic structure from such calculations is challenging in this complex, multi-band system. Our minimal model instead provides a simplified description of the relevant physics.

For comparison with experimental data, we calculate the QPI patterns using the continuum LDOS (cLDOS) method[41,42] (see methods). The calculations reveal distinct QPI patterns for the SOC-induced VHs, showing excellent agreement with the experimental data. The calculated QPI map is shown in the right half of Fig. 3b, f, g, where we clearly observe features related to the VHs that is situated just below the Fermi energy. While we do not expect the model to yield perfect agreement with experiment, given that it neglects several bands and, e.g., details of the properties of the scatterer and contributions from other layers, the agreement of the features due to the VHs around the atomic peaks (Fig. 3f, g) is excellent and most experimentally observed scattering vectors are captured (Fig. 3b). Taken together, our spectroscopic measurements and model calculations thus allow us to identify that the VHs in the immediate vicinity of the Fermi energy arises as a result of a spin-orbit coupling induced gap in the band structure, rather than one of the VHss at the zone face of the underpinning non-relativistic band structure. A similar spin-orbit coupling induced gap is found in the surface layer of $Sr_2RuO_4$[43,44], where octahedral rotations, similar to those in the bulk of $Sr_4Ru_3O_{10}$, arise as a consequence of a surface reconstruction[45].

While the QPI data is very well described by a model of the band structure accounting only for a single layer of $Sr_2RuO_4$, such a minimal model (dark blue lines in Figs. 3e and 4a) only captures approximately half of the features seen in ARPES. This difference can be ascribed to the different probing depths of the two methods: while STM is almost only sensitive to the electronic states in the top-most layer, ARPES can probe electronic states arising within the first few layers, and thus can be expected to be sensitive to the trilayer-split electronic states expected for the multi-layer structure of $Sr_4Ru_3O_{10}$. In fact, we find a generally good agreement with both the extracted dispersions (Fig. 4a) and our measured Fermi surfaces (Fig. 3e) if we take two copies of our minimal model, one as fit to the QPI data (dark blue lines), and another shifted by 40 meV (light blue lines) to account for the different octahedral rotation and splitting due to interlayer hopping. This reproduces the hole-like states with the smallest $k_F$ around the M-point associated with another saddle-point VHs located above $E_F$ here, and also well reproduces the flat bands observed near Γ.

## Discussion

From the ARPES and QPI data and comparison with the minimal tight-binding model for the band structure, we have shown that—from a hierarchy of VHss that are formed in the low-energy electronic structure of $Sr_4Ru_3O_{10}$—the most relevant is a saddle point VHs. This is created by spin-orbit coupling, resulting in the hybridization of a light spin-minority band with a heavy spin-majority band (Supplementary Fig. 8), and generating a VHs which lies below and extremely close to the Fermi energy. Given its proximity to $E_F$, this is a prime candidate for driving the metamagnetic behavior of $Sr_4Ru_3O_{10}$.

How a VHs influences macroscopic properties, however, depends on the form of the density of states divergence associated with the VHs, which in turn is known to depend sensitively on the details of the band dispersion and its symmetry[2,17,46]. A full understanding of the origin of the VHs allows determining its impact on the density of states and potentially thermodynamic properties. From our study of the low-energy electronic structure by ARPES and QPI, we can deduce that the VHs closest to the Fermi energy is a saddle point VHs with two-fold symmetry originating from a hybridization gap induced by the interplay of spin-orbit coupling and the octahedral rotations. This is different from the case of $Sr_3Ru_2O_7$, where it has been suggested that the VHs at the zone corner acquires fourfold symmetry and is of higher order due to the Brillouin zone folding[17]: here the VHs remains unaffected by the folding.

Our detailed understanding of the VHs here furthermore allows us to explore the effect of an applied magnetic field on the electronic structure using our minimal model. For an out-of-plane magnetic field, because the VHs is due to hybridization between a spin-majority and a spin-minority band, its energy shift with magnetic field does not behave simply as might be expected for a Zeeman-like behavior, but is dominated by how the crossing point between the two bands changes in the field. This can result in apparent g-factors which acquire any value between zero and infinity even in the absence of correlation effects, depending on the relative slope of the two bands which hybridize. Here, due to the heavier nature of the spin-majority band, the VHs moves away from the Fermi energy with magnetic field, as seen experimentally in Fig. 3i–k.

The situation is different for an in-plane magnetic field, where a rich phase diagram is known to result (Fig. 1b). We can model this by considering the effect of the magnetization rotating away from the c-axis. We extract the magnetization direction from measurements performed in a field at an angle of 81.6° from the c-axis (Fig. 4b, data reproduced from ref. 47). In zero magnetic field, the magnetization is along the c-axis, whereas an in-plane magnetic field tilts the magnetization from out-of-plane towards the in-plane direction by an angle θ. The metamagnetic transition in $Sr_4Ru_3O_{10}$, detected in the magnetization, occurs when the magnetization **M** exhibits an angle of ~20° with respect to the c-axis, as indicated by the red color in Fig. 4b.

We introduce the magnetization direction $\mathbf{M}_0$ into our minimal model, resulting in the Hamiltonian[48,49]

$$H = H_0 + \sum_i I \mathbf{M}_0 \mathbf{S}_i + \sum_i \lambda \mathbf{L}_i \mathbf{S}_i, \tag{1}$$

where $H_0$ is the nonmagnetic tight-binding model, $I$ the exchange splitting, $\mathbf{M}_0$ a vector of unit length pointing in the magnetization direction and $\lambda$ the spin-orbit coupling constant. Figure 4c shows the resulting change of the band structure between out-of-plane (θ = 0°) and in-plane (θ = 90°) direction of the magnetization. With increasing angle θ, the VHs at the top of the spin-orbit coupling induced gap is pushed across the Fermi energy. This change in energy is a direct consequence of the spin-orbit coupling term, $\lambda \mathbf{L} \cdot \mathbf{S}$, and the orbital character of the $d_{xz}/d_{yz}$ VHs at the M point. With rotation of the magnetization direction towards the in-plane direction, the VHs gets pushed towards the Fermi energy and as a result the upper edge of the hybridization gap crosses $E_F$. For **M**||[001], SOC results in hybridization between the minority spin VHs of $d_{xy}$ character above $E_F$ and the majority-spin $d_{xz/yz}$ VHs, pushing the latter down. When the magnetization rotates towards the [110]-direction, this hybridization is reduced until it vanishes, while SOC now results in hybridization between the two $d_{xz/yz}$ VHss of opposite spin character at the M-point, pushing the upper one towards the Fermi energy (see Supplementary Fig. 9). The spin-orbit coupling induced gap shifts with the $d_{xz}/d_{yz}$ VHs as the magnetization is rotated. For the parameters of the tight-binding model determined by fitting the QPI dispersions, the VHs already traverses the Fermi energy for θ ~ 15° between the magnetization and the c-axis. Such a scenario is also consistent with specific heat data, which show a peak in the temperature dependence that moves towards lower temperatures with increasing in-plane field[50], suggesting a VHs approaching the Fermi energy. The resulting Lifshitz transition and accompanying divergence of the density of states might provide the trigger for the metamagnetic phase transition observed in the magnetization and thermodynamic quantities. The tight binding model developed here can serve as a starting point for more sophisticated methods to describe the metamagnetic transition that captures electronic correlation effects. We expect that a similar mechanism is relevant for the field-induced Lifshitz transition observed in the surface layer of $Sr_3Ru_2O_7$[35].

From ARPES and QPI measurements, we have detected multiple VHss in the low-energy electronic structure of $Sr_4Ru_3O_{10}$. The combination of the two techniques enables us to map out the nature, symmetry, and spin character of the VHs that is closest to the Fermi energy. Our data suggests that $Sr_4Ru_3O_{10}$ is on the verge of a Lifshitz transition, similar to what has been proposed for $Sr_3Ru_2O_7$, however here the transition is not triggered by out-of-plane fields due to the dominant spin-majority character of the VHs. Our results highlight the role of spin-orbit coupling-induced hybridization gaps and VHss for the metamagnetic properties of $Sr_4Ru_3O_{10}$. An important role of VHss was previously proposed from magnetization measurements[27], however lacking spectroscopic evidence. We expect that our results, with further studies of the in-plane field-dependence of the VHss identified here, will enable a full microscopic understanding of metamagnetism in $Sr_4Ru_3O_{10}$.

## Methods
### Single crystal growth
Single crystals of $Sr_4Ru_3O_{10}$ were grown by the floating-zone method with Ru self-flux as described in ref. 51. The feed rods were prepared by a standard solid-state reaction subjecting mixed $SrCO_3$ (99.99%) and $RuO_2$ (99.9% purity) to repeated thermal cycles. An amount of excess $RuO_2$ was added to the starting powders to compensate the evaporation of Ru from the melting zone during the growth. To assess the sample quality, x-ray diffraction, energy and wavelength dispersive spectroscopy as well as electron backscattered diffraction were performed. The samples for the experiments were shaped in small rectangular pieces with average size of $1 \times 1 \times 0.3$ mm$^3$. The sample used in this work was from the same batch as the samples used in ref. 29.

### Angle-resolved photoemission
ARPES measurements were performed at the Bloch beamline of the MAX IV synchrotron. Measurements were performed with photon energies ranging from 21 eV to 67 eV using both linear horizontal and linear vertical polarized light. The probing spot size at Bloch is ~10 × 15 µm² allowing for sample regions of greatest uniformity and quality to be probed. Samples were mounted on a six-axis manipulator and cooled to 20 K. Samples were cleaved in-situ at base temperature, and measured using a Scienta DA30 electron analyzer with a vertical analyzer slit. The angular resolution was 0.2° with an energy resolution of ≈8 meV.

### Scanning tunneling microscopy
Scanning tunneling microscopy measurements were performed using a home-built microscope operating in a dilution refrigerator at temperatures below 100 mK[52]. All measurements shown in this manuscript have been acquired at T = 75 mK unless stated otherwise. The bias voltage is applied to the sample, with the tip at virtual ground. The differential conductance has been measured using a lock-in technique, applying a modulation to the bias voltage and detecting the dI/dV signal in the current ($f_L$ = 397 Hz). The amplitude of the lock-in modulation $V_L$ used for the measurements is provided in the figure captions. Samples were cleaved in-situ at low temperatures and directly inserted into the STM head. Details of data processing of QPI data are shown in Supplementary Notes 6 and 11.

### DFT calculations and minimal model
For the minimal model reproducing our quasiparticle interference data, we use a tight-binding model derived from a Density Functional Theory calculation of a monolayer of $Sr_2RuO_4$, with 15 Å of vacuum. This model consists of two Ru atoms per unit cell and has an 11° rigid octahedral rotation between them, similar to the model presented in ref. 40. The DFT calculations were performed using Quantum Espresso[53] on an 8 × 8 × 1 **k**-grid with a wavefunction cutoff of 90 Ry and a charge density cutoff of 720 Ry. We used the Perdew–Burke–Ernzerhof exchange correlation functional. The tight-binding model was generated by projecting the Ru $4d_{xz}$, $4d_{yz}$ and $4d_{xy}$ weight from the DFT calculation onto an orthonormal basis using a modified version of Wannier90[54] to preserve the relative sign of the localized wave functions. The disentanglement was performed within a window of [−2.5,0.6] eV relative to the Fermi level, including a frozen window of [−1.45,0.33] eV. For the minimal model, the DFT-derived tight-binding model is first symmetrized, we then introduce an exchange splitting of 0.44 eV for the $d_{xz}/d_{yz}$ bands and 0.89 eV for the $d_{xy}$ band. We add a local spin-orbit coupling term with λ = 200 meV. Because in correlated materials DFT does not always capture the relative sizes of Fermi surface pockets correctly (as, for example, for the δ-pocket in the surface of $Sr_2RuO_4$[55]), the chemical potential is adjusted by 120 meV for the $d_{xz}/d_{yz}$ bands and by 102 meV for the $d_{xy}$ band. Finally, all bands were renormalized by a factor of 4 to match the experimental dispersion.

### Continuum LDOS calculations

We model the QPI using the continuum Green's function method[40–42]. From the tight-binding model introduced in the DFT calculations methods section, we calculate the momentum space lattice Green's function of the unperturbed host,

$$G_{0,\sigma}(\mathbf{k},\omega) = \sum_n \frac{\xi_{n\sigma}^\dagger(\mathbf{k})\xi_{n\sigma}(\mathbf{k})}{\omega - E_{n\sigma}(\mathbf{k}) + i\eta}, \quad (2)$$

where $\mathbf{k}, \omega$ define the momentum and energy, $\xi_{n\sigma}(\mathbf{k})$ and $E_{n\sigma}(\mathbf{k})$ are the eigenvectors and eigenvalues of the tight-binding model with band index $n$ and spin $\sigma$, and $\eta$ is an energy broadening parameter. We then Fourier transform $G_{0,\sigma}(\mathbf{k},\omega)$ to obtain the unperturbed real space lattice Green's function, $G_{0,\sigma}(\mathbf{R},\omega)$, and follow the usual $T$-matrix formalism to obtain the Green's function of the system including an impurity from

$$G_\sigma(\mathbf{R},\mathbf{R}',\omega) = G_{0,\sigma}(\mathbf{R}-\mathbf{R}',\omega) + G_{0,\sigma}(\mathbf{R},\omega)T_\sigma(\omega)G_{0,\sigma}(-\mathbf{R}',\omega), \quad (3)$$

where the $T_\sigma$-matrix

$$T_\sigma = \frac{V_\sigma}{\mathbb{1} - V_\sigma G_{0,\sigma}(0,\omega)} \quad (4)$$

describes the scattering at the impurity. Here, we consider a point-like defect with equal scattering strength in the spin-up and spin-down channel, such that $V_\sigma = V_0 \mathbb{1}$.

To realistically model the QPI such that we can compare with experiment, we use the continuum Green's function approach[40–42], which defines the Green's function in terms of the continuous spatial variable $\mathbf{r}$ as

$$G_\sigma(\mathbf{r},\mathbf{r}',\omega) = \sum_{\mathbf{R},\mathbf{R}',\mu,\nu} G_\sigma^{\mu,\nu}(\mathbf{R},\mathbf{R}',\omega) w_{\mathbf{R},\mu}(\mathbf{r}) w_{\mathbf{R}',\nu}(\mathbf{r}'), \quad (5)$$

where $w_{\mathbf{R},\nu}(\mathbf{r})$ are the Wannier functions connecting the continuum and lattice space. The quasiparticle interference is then obtained from

$$\rho_\sigma(\mathbf{r},\omega) = -\frac{1}{\pi} \text{Im} G_\sigma(\mathbf{r},\mathbf{r},\omega). \quad (6)$$

For the calculations shown here, the tight-binding model and the Wannier functions are obtained from DFT calculations discussed in the methods section above. We performed the Fourier transform of the lattice Greens function over a 2048 × 2048 $k$-grid, with an energy broadening of $\eta$ = 50 μeV. An impurity potential of $V$ = 0.5 eV was used and the real space local density of states in Eq. (6) was simulated for 128 × 128 unit cells with the impurity in the center, and with 4 pixels per unit cell. QPI calculations were done using the St Andrews calcqpi code[29,49]. The resulting $\rho(\mathbf{r},\omega)$ map was Fourier transformed to simulate the QPI map, $\tilde{\rho}(\mathbf{q},\omega)$. In the experimental data, the QPI maps contain contributions from the scattering from defects on the two Ru sites with opposite octahedral rotations. To account for this in the calculations, we average over maps calculated with the defect positioned in either site to obtain maps as shown in Fig. 3.

### Data availability
Underpinning data will be made available at ref. 56.

## Acknowledgements
C.A.M., M.N., P.A.E.M., P.D.C.K. and P.W. gratefully acknowledge funding from the Engineering and Physical Sciences Research Council through EP/R031924/1, EP/S005005/1 and EP/T02108X/1, G.R.S. and P.D.C.K. from the European Research Council (through the QUESTDO project, 714193), I.B. and S.B. through the International Max Planck Research School for Chemistry and Physics of Quantum Materials, and LCR from a fellowship from the Royal Commission of the Exhibition of 1851. R.F., R.A., M.L., and A.V. thank the EU's Horizon 2020 research and innovation program under Grant Agreement No. 964398 (SUPERGATE). MH and JC thank the Swiss National Science Foundation for support. We gratefully acknowledge MAX IV Laboratory for time on the Bloch beamline under Proposal Nos. 20210763 and 20210783, which contributed to the results presented here. This work used computational resources of the Cirrus UK National Tier-2 HPC Service at EPCC funded by the University of Edinburgh and EPSRC (EP/P020267/1) and of the High-Performance Computing cluster Kennedy of the University of St Andrews. We also gratefully acknowledge Diamond Light Source (Proposal No. SI28412) and the Swiss Light Source (Proposal No. 20181951) where some preliminary data was obtained, and we thank Matthew D. Watson for assistance.

## Author contributions
C.A.M. and W.O. performed STM measurements and analyzed the STM data, with additional supporting measurements by I.B. and M.N. P.A.E.M., B.E., G.R.S., S.B., E.A.M., and P.D.C.K. performed the ARPES measurements, which were analyzed by P.A.E.M. C.P., and M.L. maintained the Bloch beamline and provided experimental support. D.H., M.H. and J.C. performed preliminary measurements. P.A.E.M. prepared the ARPES-related figures. I.B. performed and analyzed the magnetization measurements. I.B. and L.C.R. performed the DFT calculations, L.C.R. led DFT modeling, performed DFT relaxations for the surface and projected the tight-binding model. C.A.M., W.O., and P.W. carried out continuum LDOS calculations. C.A.M. prepared the STM-related figures. R.F., R.A., M.L., V.G., and A.V. grew and characterized the samples. C.A.M., P.A.E.M., P.D.C.K., and P.W. wrote the manuscript. P.W. and P.D.C.K. initiated and supervised the project.

## Competing interests
The authors declare no competing interests.

## Additional information
**Supplementary information** The online version contains

supplementary material available at
https://doi.org/10.1038/s41535-024-00645-3.

**Correspondence** and requests for materials should be addressed to Carolina A. Marques or Phil D. C. King.

**Reprints and permissions information** is available at
http://www.nature.com/reprints

**Publisher's note** Springer Nature remains neutral with regard to jurisdictional claims in published maps and institutional affiliations.